\documentclass[12pt]{article}

\usepackage{fancyhdr}
\usepackage{color}
\usepackage{amsmath}
\usepackage{amsfonts}
\usepackage{latexsym}
\usepackage{epsfig}
\usepackage{verbatim}

\RequirePackage[text={165mm,222mm},centering,includefoot]{geometry}
\RequirePackage{layout}

\newcommand{\TR}{^{\rm T}}

\def\vom{\vec\Omega}

\date{}

\begin{document}

\newcounter{INDEX}
\setcounter{section}{0}
\setcounter{subsection}{0}
\setcounter{equation}{0}

\title{
The invariant polarisation--tensor field for deuterons in storage rings and  
the Bloch equation for the  polarisation-tensor density
%influence of noise and damping on the polarisation tensor
}

\author{ D.~P.~Barber
\\
\small{Deutsches Elektronen--Synchrotron, DESY, 22607 Hamburg,~Germany} }

\maketitle

\begin{abstract} 
\noindent
I extend and update earlier work, summarised in \cite{spin2008}, whereby
the invariant polarisation--tensor field (ITF) for deuterons in storage rings 
was introduced to complement the invariant spin field (ISF).  
Taken together, the ITF and the ISF provide a
definition of the equilibrium spin density--matrix field which, in
turn, offers a clean framework for describing equilibrium spin-1
ensembles in storage rings.  I show how to construct the ITF by
stroboscopic averaging, I give examples, I discuss  adiabatic invariance 
and I introduce a formalism for describing
the effect of noise and damping.
\end{abstract}

%\pagestyle{myheadings}
%\markright{ }

\newpage

\tableofcontents

\newpage

\section[Introduction]{Introduction
%\protect \footnotemark
}
\setcounter{equation}{0} 
This paper
explores the concept of the invariant polarisation--tensor field for
charged spin-1 particles such as deuterons in storage rings.

Spin motion for charged particles moving in electric and magnetic fields is governed by the T--BMT
equation \cite{jackson}. This describes the rate of precession
of the rest--frame, pure--state, spin expectation value $\vec S$ (``the
spin'') of a particle and  in general, the
independent variable is the time. In circular particle accelerators
and storage rings the electric and magnetic guide fields are fixed in
space so that it is more convenient to take the distance $s$ around the
ring as the independent variable \cite{bhr1}. 
The motion of a
particle is governed by the Lorentz force \cite{jackson} and 
I describe particle motion in the 6--dimensional phase space in terms of the position $u$ relative to the 6--dimensional closed (periodic) orbit.
Then, at the position $s$
and the point $u$  in phase space, 
I write the T--BMT equation as $d \vec{S}/{d s}= \vom (u(s); s) \times \vec S$ ~where  the vector $\vom (u(s); s)$ describing the 
precession axis and the rate of precession,
depends on the electric and magnetic fields in the laboratory,
and on the reference energy of the ring.
Thus both
the motion of the particle and the motion of the spin expectation
value can be treated classically.  Nevertheless, as we shall see, for
some aspects of spin motion we still need to look at the quantum
mechanics. 
For this I exploit the spin density matrix.

Earlier works have emphasised the utility of the ``{\em invariant spin field}''
(ISF) for describing equilibrium spin distributions for beams of
spin-1/2 particles and have shown how the amplitude dependent spin tune
(ADST) can be exploited \cite{dk73,hh96,hvb99,gh2006,mv2000,beh2004,behresponse}.  In
this paper I embrace spin-1 particles too by introducing the concept
of the invariant field of the Cartesian polarisation tensor, which I
call the ``{\em invariant tensor field}'' (ITF). Then I show how to define
``{\em equilibrium spin density--matrix fields}'' (EDMF) 
{\footnote {In \cite{spin2008} the longer acronym, ESDeMF, was used.}}
in terms of the ISF and ITF
and explain how an EDMF can be diagonalised by rotations of the
coordinate system when a certain ansatz for the ITF is valid. I
continue the discussion by giving examples of ensembles with and without EDMFs 
and I then address the matter of adiabatic invariants. In the process I 
illustrate the advantages of calculating in reference frames of the  
``{\em invariant frame field}'' (IFF).
After mentioning other representations for the spin-1 density matrix,
I round the paper off  by extending earlier work on the
influence  of noise and damping on the evolution of the so--called
vector--polarisation density to include their effect on the
tensor--polarisation density.

Of course, the mathematics surrounding the density matrix for spin-1 
particles was established decades ago \cite{el2001,pgs2011,MC1970} but this work shows how to embed that mathematics 
within modern structures for systematising spin motion in storage rings.

I begin by reviewing the concept of the ISF for spin-1/2 particles.

\setcounter{equation}{0}
\section {Equilibrium spin distributions for spin-1/2 particles}
The properties of a mixed spin state for spin-1/2 particles are completely
defined by the $2\times 2$ density matrix $\rho$ \cite{el2001}. The
density matrix is hermitian and its trace is constrained to a definite value (in fact unity) and it is 
therefore defined by just three
real parameters which can be chosen as the three components of the
vector polarisation  $\vec P$.  Then I write
\begin{eqnarray}
\rho = \frac{1}{2}\lbrace I_{2 \times 2} + \vec P \cdot \vec \sigma \rbrace \; ,
\label{eq:1.1}
\end{eqnarray}
where the $\vec \sigma$ is the matrix--valued 3--vector formed from the Pauli matrices 
\begin{equation}
\sigma_1 = \left( \begin{array}{cc}0 & \! -i\\ i & \,\, 0\end{array}\right) ,  \; \; 
\sigma_2 =  \left( \begin{array}{cc}1 & 0\\ 0 & \!\! -1 \end{array}\right), \;  \;   
\sigma_3 = \left( \begin{array}{cc}0 & \,\, 1 \\ 1 & \,\, 0 \end{array}\right) \, ,
\label{eq:2}
\end{equation}
representing the normalised spin operators $\hat s_i ~~(i=1,2,3)$ and $I$ is the
unit matrix.  The labels $1$, $2$ and $3$ are chosen to be consistent
with those in \cite{mv2000, njp2006} and refer to the axes of a
right--handed Frenet--Serret coordinate system attached to the design
orbit whereby in the arcs of a flat ring, axis 2 is vertical, axis 1
is radial and axis 3 is longitudinal.  For spin-1/2 particles we
normalise the length of $\vec S$ to unity.  The polarisation vector is
the mixed--state expectation value of the normalised spin operator  and
it can be written as $\vec P = {\rm Tr}(\rho \vec \sigma)$.  Its components are
 $P_i = \langle \hat s_i \rangle$ where the
brackets $\langle \rangle$ signal computing the expectation value.
{\color {black}
I now follow the argumentation in \cite[Section 8.6.1] {behII}
whereby in accelerators the statistical properties of a spin-orbit
system can be expressed in terms of a spin-orbit Wigner function
\cite{mont98}  and I then write a density matrix as a function of
classical variables, namely the position $u$ in phase space and the
distance $s$ around the ring.  In Sections 2, 3 and 4, the classical orbital
motion is defined by a Hamiltonian so that the density in phase
space is conserved along a particle trajectory. This allows us to focus our
attention on the spin density matrix which I write in terms of the  
local vector polarisation ${\vec P}_{\rm loc}(u; s)$ \cite{mont98,dbkh98,beh2004,behresponse} as 
$\rho = \frac{1}{2}\lbrace I + {\vec P}_{\rm loc}(u; s) \cdot \vec \sigma
\rbrace$. }
The degree of
polarisation at a point in phase space, ${P}_{\rm loc} =  |{\vec
  P}_{\rm loc}|$, is at most unity.  Of course, since the density of
particles in phase space is finite, the notion of a mixed spin state
at each position $(u; s)$ is an idealisation, but this and other
idealisations in this paper will not detract from the value of the
basic concepts presented. In Section 6 I show how to proceed when the
phase space density is not preserved along a trajectory.

\subsection{The invariant spin field}
Since the T--BMT equation is linear in $\vec S$ and since the
particles at $(u; s)$ all see the same ${\vec{\Omega}} (u; s)$,
~ ${\vec P}_{\rm loc}(u(s); s)$ obeys the T--BMT
equation along trajectories. Of course, the same result emerges by exploiting the equation of motion for $\rho$ 
\cite[Section 1--8c]{roman}, \cite{mont98}.  Furthermore, ${P}_{\rm loc}$ is constant along 
a trajectory. For a storage ring at
fixed energy, $\vec {\Omega}$ is 1--turn periodic in $s$ at a fixed
position in phase space $u$ so that ${\vec {\Omega}}(u; s) =
{\vec {\Omega}}(u; s +C)$ where $C$ is the circumference.  This
opens the possibility of a configuration for the polarisation 
that is the same from turn to turn in the sense that
${\vec P}_{\rm loc}(u; s)$ is 1--turn periodic in $s$ for fixed $u$,
i.e., ${\vec P}_{\rm loc}(u; s + C) = {\vec P}_{\rm loc}(u; s)$. For reasons
that will become clear I denote such a ${\vec P}_{\rm loc}$ by ${\vec P}_{\rm eq}$. 
Since ${\vec P}_{\rm eq}$ also obeys the T--BMT equation 
we then have
\begin{eqnarray}
{{\vec P}_{\rm eq}}({M}(u; s+C, s); s + C) = {{\vec P}_{\rm eq}}({M}(u; s+C, s); s) = R(u; s+C, s) {{\vec P}_{\rm eq}}(u; s) \; , 
\label{eq:0}
\end{eqnarray}
where ${M}(u; s+C, s)$ is the new position in phase space
after one turn starting at $u$ and $s$, $R(u; s+C,s)$ is the
corresponding $3 \times 3$ spin transfer matrix representing the
solution to the T--BMT equation for one turn from $s$ to $s+C$ and where
here, ${\vec P}_{\rm eq}$ is represented by a column vector of its
components. By writing the T--BMT equation in matrix form we have 
\begin{eqnarray}
 \frac{d}{ds} R(u(s); s, s_0) =  {\tilde{\Omega}}(u(s); s) R(u(s); s, s_0)\; , 
\label{eq:0.4}
\end{eqnarray}
where 
\begin{eqnarray}
\tilde{\Omega} = \left( \begin{array}{ccc}0 &  -{\vec \Omega}_3 & \,\,\,\, {\vec \Omega}_2 \\
\,\,\,{\vec \Omega}_3 & \,\,\,0 &  -{\, \vec \Omega}_1 \\ 
\! -{\vec \Omega}_2 & \,\,\,\,\,\, \,{\vec \Omega}_1 & \,\,\,\,0 \end{array}\right)  \; .
\label{eq:0.5}
\end{eqnarray}
{\color{black} Of course, since it represents a pure rotation of a
$3$-vector, $R$ is $SO(3)$-valued, i.e., $R\TR R=I_{3\times 3}$ and
$\det(R)=1$. Also, note that the fact that $R\in SO(3)$ follows easily from the fact that $\tilde \Omega$ is antisymmetric \cite{magnus}.}

The relations (\ref{eq:0}) 
motivate the introduction of a
vector {\em field} ${\hat n}(u; s)$ of fixed, and in particular, unit length, obeying
similar constraints, namely
\begin{eqnarray}
{\hat n}({M}(u; s+C, s); s + C) = {\hat n}({M}(u; s+C, s); s) = R(u; s+C, s) {\hat n}(u; s) \; , 
\label{eq:0.1}
\end{eqnarray}
where $\hat n$ is represented by a column vector of its components.
Where appropriate, a similar convention will be used in the rest of
this paper.  When it exists, the field $\hat n$ is simply a 3--vector
function of $u$ and $s$ obeying the T--BMT equation and the
periodicity conditions in (\ref{eq:0.1}). No reference to real
particles and their spin states need be made. Since the vector field
${\hat n}(u; s)$ is invariant from turn to turn and independent
of the real state of a beam it is called the invariant spin field
(ISF). The ISF can be used to define the amplitude dependent spin tune
(ADST) and together they provide a most elegant way to systematise spin
motion in storage rings and circular accelerators
\cite{dk73,hh96,hvb99,gh2006,mv2000,beh2004,behresponse,ky86}.  Note that if all parameters
of the system, such as the energy, are fixed, the scalar product
$I_{\rm sn} = \vec S \cdot \hat n$ is invariant along a particle trajectory,
since both vectors obey the T--BMT equation. Thus  the
motion of $\vec S$ is simply a precession around the local
$\hat n$.

On the closed orbit, $u = 0$ and I denote the vector
$\hat n$ on the closed orbit by $\hat n_0 (s) := \hat n(0; s)$.
The vector $\hat n_0 (s)$ is 1--turn periodic and is given by the real,
unit--length eigenvector of $R(0; s+C, s)$.

For the remainder of the paper I assume that the orbital motion is
integrable to a good approximation so that $u$ can be parametrised in
terms of three pairs of action--angle variables $(J_i,
\phi_i,~i=1,2,3)$ which I abbreviate by $(J,\phi)$.  
I will use the symbols $u$ and $(J,\phi)$ interchangeably.
The actions $J$ are 
constants of the motion. Thus the orbital
phase space is partitioned into disjoint tori, each of which is
characterised by a unique set $J$. I also assume that the orbital
motion is nonresonant so that, in time, a trajectory covers its
torus. {\color{black} In \cite{behII} this is referred to as topological transitivity.} Then from (\ref{eq:0}) the norm, $P_{\rm eq}$, of ${\vec P}_{\rm eq}$ must be the same 
at all points $\phi$ on a torus but it can depend on the $J$. Furthermore
I will consider only ISFs which are continuous in $\phi$ and I avoid 
spin--orbit resonances \cite{beh2004,behresponse}. Then apart from a global  sign, $\hat n(J,\phi; s)$ is a unique 
function of $\phi$ and $s$, $2 \pi$--periodic in $\phi$ \cite{beh2004,behresponse}. So  I may 
write ${\vec P}_{\rm eq}(J,\phi; s) = {P}_{\rm eq}(J) \, \hat
n(J,\phi; s)$, bearing in mind that $|\hat n| = 1$. The requirement that $\hat n$ be
continuous in $\phi$ is chosen to reflect the expectation that the polarisation ${\vec P}_{\rm eq}$  in a real
ring varies continuously in $\phi$, and it corresponds to the requirement in \cite{behII}
{\footnote {\color{black} Note that, for most rational orbital tunes an $\hat n$  
can be extracted trivially as the normalised real
eigenvector of a multi--turn spin map \cite{njp2006}. However, for the subset of rational tunes
corresponding to so-called ``snake resonance''
the vector  obtained in this way is discontinuous in $\phi$ and it then does not
match our requirements.  Moreover, an
infinite number of totally discontinuous invariant fields satisfying (\ref{eq:0.1}) can easily be
envisaged through the so--called ``filling--up method''
\cite{eh2007}.}}.

\vspace{3mm}

The key aspects of the ISF and the ADST can be summarised as follows. 
\begin{itemize}
  
\item [1.]  
For a turn--to--turn invariant particle distribution in
  phase space, a distribution of spins in which each is initially
  aligned parallel to the ISF at its position in phase space, remains
  invariant (i.e., in equilibrium) from turn to turn, and the ISF gives
  the direction of the equilibrium polarisation ${\vec P}_{\rm eq}$ at each $(u; s)$.
{\color {black} Of course,  the polarisation for the whole torus, defined as the average
${P}_{\rm eq}(J)/({2 \pi})^3 \int^{2 \pi}_0 \int^{2 \pi}_0 \int^{2 \pi}_0 \hat
n(J,\phi; s) d \phi$, 
is invariant from turn to turn.
An example in which the polarisation for the torus is not invariant from turn to turn is presented in 
\cite{hh96, gh2006}. }

\item [2.]
For integrable orbital motion
and away from orbital resonances and spin--orbit resonances \cite{beh2004,behresponse}
the ISF determines both the maximum attainable
time-averaged polarisation and the maximum equilibrium polarisation, $P_{_{\rm
    lim}} = |\langle\hat n(J,\phi; s)\rangle_{\phi}|$, on a phase space torus
at each $s$, where the brackets $\langle\rangle_{\phi}$ denote the average over the
orbital phases. 
\item [3.]  
Away from orbital resonances and spin-orbit resonances,
$I_{\rm sn}$ is an adiabatic invariant along a particle trajectory, i.e., it does not change as a  parameter such
as the reference energy is slowly varied. See \cite{hde2006} for details. 
So, if a $\vec S$ at some $(u; s)$ is set parallel to the $\hat n (u; s)$ and then propagated forwards
with $R$ as the  parameters are 
slowly varied, the $\vec S$, which is changing {\em dynamically} finds itself parallel 
to the {\em pre-calculated} $\hat n(u';s')$  at each new position $(u';s')$ along the trajectory.

\item [4.]
The ISF provides the main axis
for orthonormal coordinate systems constructed at each point in phase space for
defining the ADST. This, in turn,  is used to define the concept
of spin--orbit resonance. Away from orbital resonance 
and spin--orbit resonance, $\hat n(u; s)$
is unique up to a global sign \cite{beh2004,behresponse}.
These coordinate systems comprise the so-called invariant frame field (IFF) mentioned later. 
\item [5.]
On the closed orbit, the ADST reduces to the number of precessions of
a spin, per turn, around $\hat n_0$. I denote this spin tune by
$\nu_0$. Its fractional part can be extracted from the complex
eigenvalues $e^{\pm 2\pi i {\nu}_0}$ of $R(\vec 0; s+C, s)$.  For a
perfectly aligned flat ring with no solenoids, $\nu_0 = a \gamma_0$
where $a$ is the gyromagnetic anomaly and $\gamma_0$ is the Lorentz
factor for the beam energy.  
\end{itemize}
These and other matters are explained and
illustrated in great detail in the sources cited above.  In order to
limit this paper to a reasonable length I will assume that the reader
is familiar with that material.

\vspace{4mm}
The most general, {\em model--independent} way, to construct the ISF is by
so--called stroboscopic averaging \cite{hh96,gh2006,mv2000,eh2007}.
This just requires a spin--orbit tracking code such as those listed in  {\color{black} \cite{mv2000} 
and \cite{Abell2015}}   
which deliver 
the  matrices $R$ along particle orbits.
As explained in \cite{hh96,gh2006,mv2000,eh2007}, with stroboscopic averaging the ISF, $\hat n({u}_0; s_0)$, 
at the starting positions ${u}_0$ and $s = s_0$
can be found in terms of multi--turn spin transfer matrices by taking the average
\begin{eqnarray}
&&  {\vec f }_N({u}_0; s_0) := \frac{1}{N+1}
\sum_{k=0}^{N} \; R(u(s_0-kC); s_0,s_0-kC)) \hat n_0(s_0) \; ,
\label{eq:1}
\end{eqnarray}
for very large $N$ and normalising this to unity: $\hat n({u}_0;
s_0) = {\vec f }_N({u}_0; s_0)/|{\vec f
  }_N({u}_0; s_0)|$. In this expression I have used
notation similar to that in \cite[eq. 22]{hh96} and chosen $\hat
n_0$ as the ``seed'' spin field, although more general choices could
be used \cite{eh2007}.  
If the orbital motion is integrable and  nonresonant, the
vector $\hat n$ need only be calculated by this means at one position
(${u}_0, s_0$). After that, $\hat n$ can be found all over the corresponding 
torus by
propagating this initial $\hat n$ along the trajectory.  Thus for integrable,
nonresonant orbital motion, there is usually {\em no} need to execute
stroboscopic averages on a grid of pre--chosen positions $\phi$
on a torus since the approach just suggested suffices for obtaining
the significant information, namely averages. 

Typical plots of $P_{_{\rm lim}}$, the ADST or the components of the
ISF, can be found in \cite{hh96,gh2006,mv2000,njp2006,hv2004}.
{\color {black} The figures in \cite{spin2010} show how to confirm that 
the $\hat n$ obtained by forward propagation is a single valued function of 
the position in phase space.} 

For detailed
discussions on convergence of stroboscopic averages, see
\cite{eh2007,hh96,gh2006}. 
{\color {black}See \cite{behII} for a framework for addressing the question of the existence of the ISF
for topologically transitive systems.}

\subsection{The equilibrium  spin density--matrix
field for spin-1/2  particles}

The existence of a 1--turn periodic ${\vec P}_{\rm eq}$ and the
corresponding ISF implies the existence of an equilibrium ~spin density--matrix
~field ~(EDMF), 
%{\color{red} ESWFF} 
~$\rho^{\rm eq}$, ~obeying ~the ~periodicity ~condition 
${\rho^{\rm eq}}(u; s + C) = {\rho^{\rm eq}}(u; s)$ and which
we can write in the form:
\begin{eqnarray}
\rho^{\rm eq}_{P_{\rm eq}(J)}(J,\phi; s) = \frac{1}{2}\lbrace I + P_{\rm eq}(J) \, \hat n(J,\phi; s) \cdot \vec \sigma \rbrace \; ,
\label{eq:1.12}
\end{eqnarray}
in an obvious notation.
Different ensembles of spins at a $(u; s)$ with the same $\vec P_{\rm eq}$
cannot be distinguished by measurements from a mixture of spins in
eigenstates of the operator $\hat n \cdot \vec \sigma$ with that $\vec
P_{\rm eq}$.  The corresponding density matrix at each $(u; s)$
is diagonal in a coordinate system in which the components of $\hat n$
are $(0,~1,~0)$.  A {\em field} of coordinate systems, namely the invariant frame field (IFF) \cite{beh2004}, provides such coordinate systems 
{\footnote {Note that in contrast to our treatment, $\hat n$ has components $(0,~0,~1)$ in the IFF of \cite{beh2004}.} }.

\section{Equilibrium  spin distributions for spin-1 particles}
\setcounter{equation}{0} 
Spin--1 particles such as deuterons have
three eigenvalues, namely $+1, 0, -1$, for the projection of the
normalised spin operator $\hat s$ onto a chosen quantisation axis, and a $3
\times 3$ density matrix. Since it is hermitian and its trace is constrained to a definite value, namely to unity, this density matrix can be completely
specified in terms of eight real parameters.  Three of these can be
the components of a vector polarisation $\vec P$ analogous to those for spin-1/2 particles
and the other five are the so--called tensor polarisations. Various
representations of the latter are in use.  For us, a  particularly useful 
parametrisation
for the density matrix $\rho$ is that given in \cite[Section 3.1.12]{el2001} in terms of 
a rank--2, $3\times3$,  real, symmetric, traceless,  Cartesian tensor $T$ as:
\begin{eqnarray}
\rho = \frac{1}{3}\left \{ I_{3 \times 3} + \frac{3}{2} \vec P \cdot {\vec {\mathfrak J}}
+\sqrt{\frac{3}{2}}\sum_{i,j} T_{ij} \ ({\mathfrak J}_i {\mathfrak J}_j +{\mathfrak J}_j {\mathfrak J}_i)\right \} \; , 
\label{eq:2.3.1}
\end{eqnarray}
where the three matrices ${\mathfrak J}$ 
\begin{equation}
{\mathfrak J}_1 = \frac{1}{\sqrt 2}\left( \begin{array}{ccc}0 & \!\! -i & 0\\ i & 0 & \!\! -i\\ 
0 & i & 0 \end{array}\right) , \; \; 
{\mathfrak J}_2 =  \left( \begin{array}{ccc}1 & 0 & 0\\ 0 & 0 & 0\\ 
0 & 0 & \!\! -1 \end{array}\right) , \;  \;   
{\mathfrak J}_3 = \frac{1}{\sqrt 2}\left( \begin{array}{ccc}0 & 1 & 0\\ 1 & 0 & 1\\ 
0 & 1 & 0 \end{array}\right) \, ,
\label{eq:2.3}
\end{equation}
representing the normalised spin operators $\hat s_i ~~(i=1,2,3)$, are the
analogues for spin-1 of the Pauli matrices and where ${\vec {\mathfrak
    J}}$ is the corresponding matrix--valued 3--vector. The tensor $T$
has just five independent components and is irreducible in that it
contains no non--zero tensors of lower rank \cite{bell75,rose}.
{\color{black} Since  ${\rm Tr}({\mathfrak J}_j {\mathfrak J}_i) = 2 \delta_{ij}$ where $\delta_{ij}$ is the Kronecker 
delta,
the tracelessness of the tensor $T$ ensures that ${\rm Tr}(\rho) = 1$, as required.}
The matrices ${\mathfrak J}$ are cyclically permuted with respect to
the corresponding matrices in \cite{el2001} because of the labeling
of the axes explained in Section 2. See the Appendix  for an informal  way to derive (\ref{eq:2.3.1}). 

The vector polarisation, with its components $P_i = \langle \hat s_i
\rangle$, is now  
{\footnote {Note that ${\rm Tr}({\mathfrak J}_i ({\mathfrak J}_j {\mathfrak J}_k + {\mathfrak J}_k {\mathfrak J}_j ) ) = 0$ $ \forall~ i,j,k $.}
$\vec P = {\rm Tr}(\rho \vec {\mathfrak J})$
and again, ${P} = |\vec P|$ is at most unity. However, 
the components of $T$ which are also needed for the spin-1 density matrix, depend on quadratic combinations of spin
operators. In particular \cite{el2001},  
\begin{eqnarray}
T_{ij} =
\frac{1}{2} \sqrt{\frac{3}{2}}\left \{\langle\hat s_i \hat s_j + \hat
  s_j \hat s_i\rangle -\frac{4}{3}\delta_{ij} \right \}  \; .
\label{eq:2.2}
\end{eqnarray}
%where the Kronecker delta has been used \cite{el2001}.  
That this is consistent with (\ref{eq:2.3.1}) can be seen  
by computing the ${\langle\hat s_i \hat s_j \rangle}$
using ${\rm Tr}(\rho {\mathfrak J}_i {\mathfrak J}_j)$. 
The degree of tensor polarisation is defined as ${\mathfrak T} := 
\sqrt{\sum_{i,j} T^2_{ij}} = \sqrt{{\rm Tr}(T T\TR)} = \sqrt{{\rm Tr}(T^2)}$ 
%It is independent of $u$ and $s$  
and it is at most unity \cite{el2001}. If ${\mathfrak T}^2 = 0$, $T = 0_{3 \times 3}$.
Another convention for defining the degree of tensor polarization is described in Section 4.

In contrast to the case of spin-1/2 particles, spin-1 particles
can exist in pure states for which $|{\vec S}|$ is zero. In that case $|{\vec S}|$ cannot
be normalised to unity \cite{manenorm}. 
%If $|{\vec S}|$ is zero $I_{\rm sn}$ vanishes and it is automatically invariant.
For the other pure states, $|{\vec S}|$ can be
normalised to unity. The overall degree of polarisation for
spin-1 particles is 
$\frac{3}{4} P^2 + {\mathfrak T}^2$  \cite[eqs. 3.1.55, ~3.1.65]{el2001}.
So an ensemble is unpolarised only when {\em both} ${P}$ and 
${\mathfrak T}$ are zero.  In that case $\rho = \frac{1}{3} I$ so that
the probabilities for the three substates are equal at 1/3.

\subsection{The invariant tensor field}
Given the existence of ISFs in most cases of interest, it is natural
to ask whether  invariant tensor {\em fields} (ITF) can exist. I now examine this
possibility by defining the ITF and then suggesting how to construct
it by stroboscopic averaging and then in terms of the ISF.

For this I use the analogue for $T$ of solutions of the T--BMT
equation for spin. 
In particular, if along a trajectory, an initial spin, ${\vec S}^{\rm
  i}$, is transformed into a final spin, ${\vec S}^{\rm f}$, by a spin
transfer matrix $R$ according to the relation
\begin{eqnarray}
\vec S^{\rm f} =  R ~\vec S^{\rm i} \; ,
\label{eq:3}
\end{eqnarray}
then the components of $T$ are transformed according to the rule \cite{bell75}
\begin{eqnarray}
T^{\rm f} =    R ~T^{\rm i} R\TR \; .
\label{eq:4}
\end{eqnarray}
Of course, $\vec P^{\rm f} =  R \, \vec P^{\rm i}$.
{\color{black}The similarity transform, (\ref{eq:4}) conserves the trace and the symmetry of the tensor 
as well as ${\mathfrak T}$.}
Moreover, 
\begin{eqnarray}
 \frac{d T}{d s}~ = ~ [\,\tilde {\Omega}, ~T] := 
{\tilde {\Omega}} \, {T} - {T} \,{\tilde {\Omega}} 
\label{eq:3.1}
\end{eqnarray}
while
\begin{eqnarray}
\frac{d \vec P}{d s}~ =~ 
{\tilde {\Omega}}\, {\vec P}  \;  .
\nonumber
\end{eqnarray}

\noindent 
I emphasise the fact, obvious from (\ref{eq:3}) and (\ref{eq:4}), that 
as soon as the matrix $R$ for rotating spins is known, the transformation
for $T$ follows trivially.
Of course, $T$ remains traceless and symmetric even 
if a parameter such as the reference  energy is varied along a trajectory since 
the $\tilde \Omega$ in (\ref{eq:3.1}) is still antisymmetric while  the $R$ in (\ref{eq:4}) therefore remains orthogonal \cite{magnus}.

\vspace{3mm}

I now define the ITF $T^{\rm I}$, in analogy with the definition for
the ISF, by the periodicity condition
\begin{eqnarray}
T^{\rm I}({M}(u; s+C, s); s + C) =
T^{\rm I}({M}(u; s+C, s); s)     = 
R(u; s+C, s) \; T^{\rm I}(u; s) R\TR(u; s+C, s)   \; , \!\!\!\!\!\!\!     \nonumber \\ 
\label{eq:5} 
\end{eqnarray}
with $\sqrt{{\rm Tr}(T^{\rm I} T^{\rm I}  )} = 1$. {\color{black} This normalisation is  
preserved along a trajectory since $(T^{\rm I})^2$ is transported as 
$ R ~(T^{\rm I})  R\TR R  ~(T^{\rm I}) R\TR = R ~(T^{\rm I})^2 R\TR$ and a similarity
transformation preserves a  trace.}
Just as with the ISF, when it exists, the ITF 
is simply a single valued tensor function of $u$
and $s$ obeying (\ref{eq:4}) and the periodicity conditions as
in (\ref{eq:5}). No reference to real particles and their spin states
need be made. Intuition suggests that the ITF is
unique up to a global sign away from orbital resonances and spin--orbit
resonances.
Stroboscopic averaging  to calculate the ISF is grounded in 
the linearity of the T-BMT equation
and the fact that ${\vec \Omega}(u; s)$ is 1--turn periodic in $s$.
The equation of motion (\ref{eq:3.1}) for $T$ is also linear. So I now {\color{black} postulate that} 
the ITF can also be constructed using stroboscopic averaging:
% in analogy with the ISF:
\begin{eqnarray} 
{ g  }_N({u}_0; s_0) :=  
\frac{1}{N+1} \sum_{k=0}^{N} R (u(s_0-kC); s_0,s_0-kC) \; T(s_0) \; R\TR (u(s_0-kC); s_0,s_0-kC) \; , \!\!\!\!\!\!\! \nonumber \\
\label{eq:6}
\end{eqnarray}
where  $N$ is very large and $T(s_0) $ is a fixed $3 \times 3$ 
symmetric matrix with zero trace.
The symmetry and tracelessness  ensure that the $3 \times 3$ matrix 
${ g }_N({u}_0; s_0)$ is traceless and symmetric. 
The ITF should be  obtained as 
$T^{\rm I}({u}_0; s_0)  = 
{ g }_N/\sqrt{{\rm Tr}( { g }^2_N)}$.

{\color {black} 
In fact numerical tests away from orbital resonances and spin-orbit resonances in a typical ring confirm that
stroboscopic averaging as in (\ref{eq:6}) does indeed deliver a tensor
$T$ with the required properties, namely those of a
$T^{\rm I}$.
In particular when one propagates this $T^{\rm I}({u}_0; s_0) $ forward turn-by-turn
to obtain $T^I$ at $s_0$ all over the 
corresponding torus, 
one finds a single valued tensor function of the position in phase space $u$, as required.
Moreover, if the stroboscopic average in (\ref{eq:6}) is calculated for many
different, non--zero,  traceless, symmetric matrices $T(s_0)$, 
then, away from orbital
resonances and spin--orbit resonances, the same normalised stroboscopic
average is obtained (up to a sign).
%and it is then again a single valued function of the orbital phases.  
This suggests, but does not prove, that the
ITF is unique up to a global sign.

Given the simplicity of the rule (\ref{eq:4}) it is also tempting to
try to express the ITF in terms of the ISF, and the form: 
\begin{eqnarray}  
 T^{\rm I}  = \pm   {\sqrt{\frac{3}{2}}} \left \{ \hat n {\hat n}\TR  - \frac{1}{3} I_{3 \times 3} \right \} \; ,
\label{eq:7}
\end{eqnarray}
suggests itself. The term with the $3 \times 3$ unit matrix ensures
that the tensor is traceless and the factor ${\sqrt{\frac{3}{2}}}$
ensures the chosen normalisation.  
In fact one finds in numerical tests that the ansatz  (\ref{eq:7}) agrees perfectly with the normalised
stroboscopic average of (\ref{eq:6}).}

Since a spin transfer matrix $R$ is
orthogonal, it is trivial that this ansatz satisfies (\ref{eq:4}).
Note that the required periodicity conditions (\ref{eq:5}) are
fulfilled owing to the analogous periodicity of $\hat n$. The ansatz
(\ref{eq:7}) clearly satisfies all requirements for $T^{\rm I}$.

Later we shall see that the ansatz (\ref{eq:7}) is supported by
another consideration based on requirements for pure states. 
In fact, as argued in \cite{bkv2009}, the ansatz (\ref{eq:7}) is unique 
if the ISF is unique and it then takes this form.
{\color {black}
The same conclusion is reached in a  broader context and in a more powerful manner in \cite[Section 8]{behII} 
whereby invariant fields are associated with symmetry groups.  }
I therefore accept the ansatz of (\ref{eq:7}) from here on.
%So we will adopt this form in the remainder of
%the paper. 
Moreover it will suffice to choose the $+$ sign.
Then, for example,  
$T^{\rm  I}_{2,2} (J,\phi; s) = \sqrt{\frac{2}{3}}(\frac{3}{2}\cos^2 \theta -
\frac{1}{2})$ where $\theta = \cos ^{-1} (\hat n _2 (J,\phi; s))$. 

Since $T^{\rm I}$ is unique, it can be obtained using
(\ref{eq:7}) as soon as the ISF is known, without further stroboscopic
averaging. Then, for example,
in a parameter regime where the so--called
single resonance model \cite{njp2006,spin2010,mane2} provides an approximation
to the spin dynamics, one not only has an approximate analytical
formula for the ISF, but also for the ITF.

\subsection{The EDMF for spin-1 particles}
The kind of argument that leads to the concept of the equilibrium
polarisation ${\vec P}_{\rm eq}(J,\phi; s) = P_{\rm eq} \, \hat n(J,\phi; s)$ on a torus, also leads to the concept of the equilibrium
polarisation tensor $T_{\rm eq}(J,\phi; s) := {{\mathfrak T}_{\rm
    eq}}(J) \, T^{\rm I}(J,\phi; s)$ where ${\mathfrak T}_{\rm eq}(J)$ is
the degree of equilibrium tensor polarisation. 
If the particles are distributed uniformly in $\phi$ on the torus $J$
and $T (J,\phi; s) = T_{\rm eq}(J,\phi; s)$ at some $s$, the polarisation tensor
for the torus, 
defined as 
${\mathfrak T}_{\rm eq}(J)/({2
  \pi})^3 \int^{2 \pi}_0 \int^{2 \pi}_0 \int^{2 \pi}_0 T^{\rm I} (J,\phi; s) d \phi$, is invariant
from turn to turn.

With the ITF we can now construct an EDMF for spin-1 particles: 
\begin{eqnarray}
\rho^{\rm eq}_{[P_{\rm eq},{\mathfrak T}_{\rm eq}](J)}(J,\phi; s) = \frac{1} {3}\left \{ I + \frac{3}{2} P_{\rm eq}(J)\hat n \cdot {\vec {\mathfrak J}}
+ {\sqrt{\frac{3}{2}} {{\mathfrak T}_{\rm eq}(J)} } \sum_{i,j} T^{\rm I}_{ij} \ ({\mathfrak J}_i {\mathfrak J}_j +{\mathfrak J}_j {\mathfrak J}_i)\right \} \; .
\label{eq:9}
\end{eqnarray}
So on a torus, the EDMF is defined by two free parameters, $P_{\rm eq}(J)$ and
$ {\mathfrak T}_{\rm eq}(J)$.
As required, ${\rm Tr} (\rho^{\rm eq}) = 1$
for any pure or mixed state.  However, for a pure state we require
that ${\rm Tr} (\rho^2) = 1$.  Then in a pure state, from
(\ref{eq:9}) and (\ref{eq:7}) and after some matrix algebra, I find the constraint
\begin{eqnarray}
1 = \frac{1}{9} \left \{3 + 
\frac{9}{2} P^2_{\rm eq} 
+  6 {{\mathfrak T}^2_{\rm eq}} \right \}
\quad \Rightarrow \quad  1  =  \frac{3}{4} P^2_{\rm eq} + {{\mathfrak T}^2_{\rm eq}}
\label{eq:11} \, ,
\end{eqnarray}
which is clearly independent of the precise values of the components
of $\hat n $ and $T^{\rm I}$. In this context it should be emphasised
that the ISF and the ITF just encode the relative sizes of their
components and that to specify actual equilibrium spin distributions on tori we need
$P_{\rm eq}(J)$ and ${\mathfrak T}_{\rm eq}(J)$ too.  
Relation (\ref{eq:11}) corresponds to the
fact that pure states are fully polarised according to the definition
in \cite[eqs. 3.1.55, ~3.1.65]{el2001}.

As we are reminded in \cite{el2001}, although the density matrix for
spin-1/2 particles can always be diagonalised by a rotation of the
coordinate system, 
this is not always the case for
spin-1 particles.  However, it is always possible for the EDMF of
spin-1 particles when the ansatz in (\ref{eq:7}) is valid since the
ITF, and thus the EDMF, will be diagonal if 
$\hat n$ has components
$(0,~1,~0)$. This can always be achieved at each $(J,\phi; s)$ by
rotating the coordinate frame to make $\hat n$ vertical in that frame, e.g.,  by
viewing the system in a reference frame of the IFF.
According to the definition in \cite{el2001}, the ISF therefore
defines the quantisation axes for the EDMF on a torus for spin-1
particles too.  In particular, different ensembles of spins at a $(J,\phi; s)$
with the same $\vec P_{\rm eq}$ and ${T}_{\rm eq}$ cannot be
distinguished by measurements from a mixture of spins in eigenstates
of the operator $\hat n \cdot \vec {\mathfrak J}$ with those $\vec P_{\rm eq}$ and
${T}_{\rm eq}$.  Of course, since all spin transformations
here are rotations according to the matrices $R$, it is clear that if
$\rho^{\rm eq}$ can be diagonalised at one point on a trajectory, it
can be diagonalised by a rotation everywhere else along the
trajectory.

A way to ensure that a  tensor field is invariant at high energy is described in Section 4.

I now illustrate these concepts with simple examples in which I
choose some special configurations of spins and then check to see if
they can be described by EDMFs.

\subsection{Examples}

\vspace {5mm}

\noindent
{\bf Example 1}

\vspace {2mm}

\noindent
Consider the case where all particles at some arbitrary $(J, \phi; s)$
are in the eigenstate of $\hat n \cdot {\mathfrak J}$ whose eigenvalue is +1 
so that $P_{\rm loc} = 1$. I write this eigenstate as $|n^{+}\rangle$.
The corresponding normalised 3-spinor is  
\begin{eqnarray}
e^{i \chi}
\left( \begin{array}{c} \frac{1 + n_2}{2}  \\ \frac{n_3 + i n_1}{\sqrt{2}} \\ 
\frac{1 - n_2}{2} \frac{n_3 + i n_1}{n_3 - i n_1} 
\end{array}\right)  = e^{i (\chi + \psi)}
\left( \begin{array}{c} \frac{1 + n_2}{2}  e^{- i \psi} \\ \frac{r}{\sqrt{2}} \\ 
 \frac{1 - n_2}{2} e^{ + i \psi} 
\end{array}\right)
 \;  ,
\label{eq:10.3}
\end{eqnarray}
where $\chi$ is an arbitrary phase, $r = \sqrt{1 - n^2_2}$, $n_1 = r \sin \psi$ and $n_3 = r \cos \psi$.
Then, by evaluating the expectation values 
$\langle n^{+}|~{\mathfrak J}_i {\mathfrak J}_j ~| n^{+} \rangle$ in 
(\ref{eq:2.2}) for $|n^{+} \rangle$, I find for the local polarisation tensor
{\footnote{Alternatively, after reordering the axes as in (\ref{eq:2.3}), one can use the convenient formalism for general pure states 
described around  \cite[eq. 3.1.79]{el2001}.}: 
\begin{eqnarray}
T_{\rm loc} = 
{\frac{1}{2}} \sqrt{\frac{3}{2}} \left \{ \hat n {\hat n}\TR  - \frac{1}{3} I \right \} \; .
\label{eq:10.6}
\end{eqnarray}
So $T_{\rm loc}(J, \phi; s)$ is proportional
to the $T^{\rm I}(J, \phi; s)$ of (\ref{eq:7}).  
The proportionality factor is $1/2$ and the same value will be obtained at any other 
$(J, \phi; s)$. Also, 
it is preserved as the ensemble moves over the torus. So there is an EDMF. In particular I have 
$T_{\rm eq}(J,\phi; s) = T_{\rm loc}(J, \phi; s) = 
T^{\rm I}(J, \phi; s)/ 2$,  i.e.,  $T_{\rm eq}(J,\phi; s) :=
{{\mathfrak T}_{\rm eq}}(J) \, T^{\rm I}(J,\phi; s)$ where ${\mathfrak
  T}_{\rm eq}(J) = 1/2$. With $P_{\rm loc} = P_{\rm eq} =1$, this value of ${\mathfrak T}_{\rm eq}(J)$ satisfies (\ref{eq:11}), as expected for a pure state. 
I can now write
\begin{eqnarray}
\rho^{\rm eq} &=& \left \{ \frac{I}{3} + \frac{1}{2} P_{\rm eq}\hat n \cdot {\vec {\mathfrak J}}
+ { {{\mathfrak T}_{\rm eq}} } \sum_{i,j} 
({\hat n}_i {\hat n}_j - \frac{1}{3} \delta_{ij})
\frac{
({\mathfrak J}_i {\mathfrak J}_j +{\mathfrak J}_j {\mathfrak J}_i)}{2}\right \}\nonumber \\
  &=& \left \{ \frac{I}{3} + \frac{1}{2} \hat n \cdot {\vec {\mathfrak J}}
+ \frac{1}{2} \sum_{i,j} 
({\hat n}_i {\hat n}_j - \frac{1}{3} \delta_{ij})
\frac{
({\mathfrak J}_i {\mathfrak J}_j +{\mathfrak J}_j {\mathfrak J}_i)}{2}\right \}\; ,
\label{eq:10.1}
\end{eqnarray}
where I have suppressed the dependences on $J, \phi$ and $s$ for
convenience. It is now clear that the ansatz (\ref{eq:7}) is
natural  if (\ref{eq:9}) is to be able to encompass
these pure states on a torus. It is easily checked that
this density matrix is idempotent as required for pure states.

Note that these states are the so--called coherent states appearing in
the discussion of the matter of defining the classicality of spin
states in \cite{giraud2008}. There, the density operator for the pure state
with a spin
expectation value $\hat {\mathfrak n}$ of unit magnitude, is given by the 
projector
\begin{eqnarray}
 |\hat {\mathfrak n} \rangle \langle \hat {\mathfrak n}|  = \left \{ \frac{I}{3} + \frac{1}{2} \hat {\mathfrak n} \cdot {\vec {\mathfrak J}}
+ \frac{1}{2} \sum_{i,j} 
({\hat {\mathfrak n}}_i {\hat {\mathfrak n}}_j - \frac{1}{3} \delta_{ij})
\frac{
({\mathfrak J}_i {\mathfrak J}_j +{\mathfrak J}_j {\mathfrak J}_i)}{2}\right \}\; .
\label{eq:10.2}
\end{eqnarray}
With the explicit representation  (\ref{eq:10.3}), the equality of the 
$3 \times 3$ matrices on each side of (\ref{eq:10.2}) is readily confirmed.

Of course, if all particles 
are in the eigenstate $|n^{-}\rangle$
of $\hat n \cdot {\mathfrak J}$ whose eigenvalue is $-1$, 
$P_{\rm eq} = -1$. However, ${\mathfrak T}_{\rm eq}$ is again 1/2.
The 3-spinor for $|n^{-}\rangle$ can be obtained by simply reversing the signs of the 
components of $\hat n$ in (\ref{eq:10.3}) to obtain
\begin{eqnarray}
e^{i (\chi + \psi)}
\left( \begin{array}{c} \frac{1 - n_2}{2}  e^{- i \psi}\\ -\frac{r}{\sqrt{2}} \\ 
 \frac{1 + n_2}{2} e^{ + i \psi} 
\end{array}\right)
 \;  .
\label{eq:10.3.1}
\end{eqnarray}

\vspace {5mm}

\noindent
{\bf Example 2}

\vspace {2mm}

\noindent
Consider the case where all particles at some $(J, \phi; s)$
are in the eigenstate $|n^{0}\rangle$ of $\hat n \cdot {\mathfrak J}$ whose eigenvalue is 0 
so that $P_{\rm loc} = 0$. 
The corresponding normalised 3-spinor is  
\begin{eqnarray}
e^{i \chi}\sqrt{\frac{1 - n^2_2}{2}}
\left( \begin{array}{c} 1  \\ 
- \frac{\sqrt{2} n_2}{n_3 - in_1 } \\ 
-\frac{n_3 + in_1} {n_3 - in_1} \end{array}\right)
=  e^{i (\chi + \psi) } {\frac{r}{\sqrt 2}}
\left( \begin{array}{c} \,\,\,\,e^ {-i \psi }   \\ 
- \frac{\sqrt{2} n_2}{r} \\ 
-e^ {+ i \psi}
\end{array}\right) \;  ,
\label{eq:10.4}
\end{eqnarray}
where $\chi$ is an arbitrary phase. Then, by evaluating the expectation values in 
(\ref{eq:2.2}) for $|n^{0} \rangle$, I find for the local polarisation tensor: 
\begin{eqnarray}
T_{\rm loc} = -
\sqrt{\frac{3}{2}} \left \{ \hat n {\hat n}\TR  - \frac{1}{3} I \right \} \; .
\label{eq:10.7}
\end{eqnarray}
So  the ansatz (\ref{eq:7}) for $T^{\rm I}$ appears again. Thus there is again an EDMF and in this case ${\mathfrak T}_{\rm eq}(J) = -1$.  Since  $P_{\rm eq} = 0$,  
(\ref{eq:11}) is satisfied as expected for a pure state. 
 
\vspace {5mm}

%\newpage

\noindent
{\bf Example 3}

\vspace {2mm}

\noindent
Although ensembles in the pure states $|n^{+}\rangle$ or
$|n^{-}\rangle$ can be described by EDMFs, this is not necessarily true for
ensembles in coherent superpositions of these states. For example,
consider the case where all particles at some $(J, \phi; s)$ are in
the pure state $|N^{+}\rangle := \{|n^{+}\rangle + |n^{-}\rangle
\}/{\sqrt {2}}$ or where all particles are in the pure state
$|N^{-}\rangle := \{|n^{+}\rangle - |n^{-}\rangle \}/{\sqrt {2}}$.
In both cases $P_{\rm loc} = 0$ and ${\mathfrak T} = 1$ so that the relation 
$1  =  \frac{3}{4} P^2 + {{\mathfrak T}^2}$ for pure states is fulfilled.
However, owing to the presence of ``cross terms'' involving $|n^{+}\rangle$ and $|n^{-}\rangle$, neither of the local
polarisation tensors, $T^+_{\rm loc}$ for $|N^{+}\rangle$ and
$T^-_{\rm loc}$ for $|N^{-}\rangle$, is proportional to the $T^{\rm
  I}$ calculated using (\ref{eq:7}). So ensembles in one or the other of these
pure states are not described by an EDMF. 

Alternatively we may say that the states $|N^{\pm}\rangle$ 
cannot be represented by a projector of the form   
(\ref{eq:10.2}). This is in contrast to the case of spin-1/2 particles. 
There, all pure states can be represented in the corresponding form namely,
$|{\mathfrak n}\rangle \langle {\mathfrak n}|  =    \frac{1}{2}\lbrace I + \hat {\mathfrak n} \cdot \vec \sigma \rbrace$, since every pure state is an eigenstate of ${\hat {\mathfrak n}} \cdot \vec \sigma$  
for some unit vector ${\hat {\mathfrak n}}$.

\vspace {5mm}
%\newpage

\noindent
{\bf Example 4}

\vspace {2mm}

\noindent
I now turn to mixed states. 
Consider the mixed state at some arbitrary $(J, \phi; s)$ 
comprised of equal proportions of the pure states
$|N^{+}\rangle$ and $|N^{-}\rangle$. The density matrix is the 
average of the density matrices for the two pure states. Then the aforementioned cross terms cancel against each other and   
%$T_{\rm loc} =  \frac{1}{2} T^+_{\rm loc} + \frac{1}{2} T^-_{\rm loc}$ 
we have
\begin{eqnarray}
T_{\rm loc} =  \frac{1}{2} T^+_{\rm loc} + \frac{1}{2} T^-_{\rm loc} = 
\frac{1}{2}\sqrt{\frac{3}{2}} \left \{ \hat n {\hat n}\TR  - \frac{1}{3} I \right \} \; ,
\label{eq:10.8}
\end{eqnarray}
so that there is an EDMF and ${\mathfrak T}_{\rm eq}(J) = 1/2$. 
Since $P_{\rm eq} = 0$, $\frac{3}{4} P^2_{\rm eq} + {{\mathfrak T}^2_{\rm eq}} < 1$
as expected in a mixed state.

This example illustrates nicely how  an EDMF can exist for a mixture of states which individually
do not have EDMFs. There is an unlimited number of such examples.

\vspace {5mm}

\noindent
{\bf Example 5}

\vspace {2mm}

\noindent
Consider now the mixed state at some arbitrary $(J, \phi; s)$ in which the $(\phi; s)$-independent  fraction
$f^{+}$ is in the state $|n^{+}\rangle$ and the remainder is in the state $|n^{-}\rangle$.
Then
\begin{eqnarray}
T_{\rm loc} = 
\frac{1}{2}\sqrt{\frac{3}{2}} \left \{ \hat n {\hat n}\TR  - \frac{1}{3} I \right \} \; ,
\label{eq:10.9}
\end{eqnarray}     
and there is an EDMF with $P_{\rm eq} = f^{+} - f^{-} = 2 f^{+} - 1$ and ${\mathfrak T}_{\rm eq}(J) = 1/2$.

{\color {black} 
With  $f^{+} = 1/2$, the EDMF is the same as that in Example 4 so that the two ensembles cannot
be distinguished by measurements.}

\vspace {5mm}

\noindent
{\bf Example 6}

\vspace {2mm}

\noindent

Consider next the mixed state at some arbitrary $(J, \phi; s)$ comprising $(\phi; s)$-independent  fractions
$f^+$, $f^-$ and $f^0$ 
of the pure states $|n^{+}\rangle$, $|n^{-}\rangle$ and $|n^{0}\rangle$.
%with $f^+ + f^- + f^0 = 1$.
As we have seen, each of these pure states has a $T_{\rm loc}(J, \phi; s)$ proportional
to the $T^{\rm I}(J, \phi; s)$ of (\ref{eq:7}) and thus an EDMF.
The overall $T_{\rm loc}(J, \phi; s)$ is the average of those for the three pure states and  we find
\begin{eqnarray}
T_{\rm loc}(J, \phi; s)  = \frac{1}{2} (1-3 f^0) T^{\rm I} (J, \phi; s) \; ,
\end{eqnarray}
so that ${\mathfrak T}_{\rm eq}(J) = 1/2 \;(1-3 f^0)$. 
The factor $(1-3 f^0)$ is a local version, ${\cal A}_{\rm loc}$, of the so-called ``{\em alignment}'', ${\cal A}$, to 
be mentioned later \cite{el2001}.
Obviously $-2 \le {\cal A}_{\rm loc} \le 1$. 
As ever, $P_{\rm eq} = f^+ - f^-$.

If $f^+ =  f^- = f^0 = 1/3$,  both $P_{\rm eq}$ and  ${\mathfrak T}_{\rm eq}(J)$ are zero.
Then that ensemble is fully unpolarised.

\vspace {5mm}

\noindent
{\bf Example 7}

\vspace {2mm}

\noindent

In the previous examples, the pure states at each $(J, \phi; s)$ are defined
in terms of the ${\hat n}( J, \phi; s)$ at those points. However, it is perhaps
more realistic to examine mixtures of pure states whose quantisation
axes are tilted away from ${\hat n}(J, \phi; s)$ and distributed uniformly
around the ${\hat n}(J, \phi; s)$. I call this a ``{\em conical}'' mixed state and I consider 
three cases.

\vspace {2mm}

\noindent
{\bf {\small Case 1}}

Consider then the configuration in which  each  particle at some arbitrary $(J, \phi; s)$ is in 
an eigenstate of $\hat q \cdot {\mathfrak J}$ whose eigenvalue is +1 where the ${\hat q}$
are unit vectors tilted at the common angle $\Theta$ from the ${\hat n}(J, \phi; s)$
and uniformly distributed around the ${\hat n}(J, \phi; s)$.  Then for the resulting mixed state
$P_{\rm loc} = \cos \Theta$ and there is an equilibrium vector-polarisation field
${\vec P}_{\rm eq}(J, \phi; s) =  \cos \Theta \,  {\hat n}(J, \phi; s)$. 
 
Since for this mixed state the $\hat q$ are uniformly distributed around their $\hat n$ vectors, 
intuition suggests that   the $T_{\rm loc}(J, \phi; s)$ for this configuration should be 
proportional to the ITF. 
%This is the case, as is easily seen by comparing $T_{\rm loc}$ and $T^{\rm I}$
%within a reference frame of the  IFF at $(J, \phi; s)$.
To test this I compare $T_{\rm loc}$ and $T^{\rm I}$ within a reference frame of the  IFF at $(J, \phi; s)$.
Recall that within sach a frame  $\hat n$ has the components $(0,~1,~0)$. Then with (\ref{eq:7}) evaluated
within the that frame, the ITF is diagonal
with components
$$(T_{11}^{\rm I})^{\rm IFF}   = (T_{33}^{\rm I})^{\rm IFF}  = - \frac{1}{2} \,(T_{22}^{\rm I})^{\rm IFF} =
- {\sqrt{ \frac{3}{2}} \, \frac{1}{3} } \; ,$$
in obvious notation 
\footnote{Note that we see explicitly here that $T^{\rm I}$ has two equal eigenvalues as 
required by the arguments in \cite{bkv2009}.}.

Next, I transform $\hat q$ into the same frame and I  write the direction cosines of $\hat q$ in that frame as 
$({\sin}\Theta \, {\sin}\Phi, \, {\cos}\Theta,  \, {\sin}\Theta \, {\cos}\Phi )$ 
where $\Phi$ defines the 
direction of the projection of a $\hat q$ onto the plane perpendicular to the $\hat n$. 
From the definition of this mixed state, $\Phi$ is distributed uniformly
between $0$ and $2\pi$.  

In analogy with the calculations 
for Example 1,   the local polarisation tensor, $T_{\rm loc}^q$, viewed in the machine coordinates, 
at some $(J, \phi; s)$ for each $\hat q$ is   
\begin{eqnarray}
T_{\rm loc}^q := \frac{1}{2}
\sqrt{\frac{3}{2}} \left \{ \hat q {\hat q}\TR  - \frac{1}{3} I \right \} \; .
\label{eq:10.10}
\end{eqnarray}

An analogous formula, in terms of the $\hat q$  viewed in the chosen frame  obtains  
and the $T_{\rm loc}^{\rm q, IFF}$ has a simple non-diagonal form in terms of the above-specified direction cosines.
Then by averaging the $T_{\rm loc}^{\rm q, IFF}$  over $\Phi$
the resulting $T_{\rm loc}^{\rm IFF}$  at the  chosen $(J, \phi; s)$ is diagonal with components 
$$T_{{\rm loc},11}^{\rm IFF}  = T_{{\rm loc},33}^{\rm IFF}  = - \frac{T_{{\rm loc},22}^{\rm IFF}}{2} = 
\frac{1}{2} \sqrt{\frac{3}{2}} \, \biggl( \frac{ {\sin}^2 \Theta}{2} - \frac{1}{3} \biggr) \, . $$
Again, the notation should require no explanation. Note that when viewed in a reference frame of the IFF, the ITF
is a constant matrix independent of $(J, \phi; s)$ and therefore trivially invariant.

With this we find: 
$$T_{{\rm loc}}^{\rm IFF} =
 - \frac{3}{2} \, \biggl( \frac{ {\sin}^2 \Theta}{2} - \frac{1}{3} \biggr) \, (T^{\rm I})^{\rm IFF}.$$

Then after transforming back to machine coordinates we have 
$$T_{\rm loc} =
 - \frac{3}{2} \, \biggl( \frac{ {\sin}^2 \Theta}{2} - \frac{1}{3} \biggr) \, T^{\rm I} \; ,$$

so that 
$${\mathfrak T}_{\rm eq} = - \frac{3}{2} \, \biggl( \frac{ {\sin}^2 \Theta}{2} - \frac{1}{3} \biggr) \, .$$
Thus the intuition is vindicated. As required ${\mathfrak T}_{\rm eq} = 1/2$ when $\Theta = 0$ or $\pi$.

Since differently constituted mixed states with the same $\rho^{\rm eq}$ cannot be distinguished by measurements,
I now check to see whether this configuration can be mimicked by a  configuration from  Example 6 with 
a special choice of $f^+, \, f^-$ and $f^0$. 

The vector polarisation can always be reproduced simply by choosing values of $f^+$ and  $f^-$ such that
$f^+ - f^- = {\cos} \,\Theta$. However, to mimic $T_{\rm loc}$ one needs 
\begin{eqnarray}
- \frac{3}{2} \, \biggl( \frac{ {\sin}^2 \Theta}{2} - \frac{1}{3} \biggr) \, T^{\rm I} = 
     \frac{1}{2} \; (1 - 3 f^0) T^{\rm I} \; .
\end{eqnarray}
so that 
\begin{eqnarray}
f^0 = \frac{ {\sin}^2 \Theta}{2} \; .
\end{eqnarray}
Then
\begin{eqnarray}
f^+ +  f^- = 1 - \frac{ {\sin}^2 \Theta}{2} \; ,
\end{eqnarray}
and 
\begin{eqnarray}
f^{\pm} = \frac{1}{2} \biggl ( 1 \, {\pm} \,{\cos} \, \Theta -   \frac{ {\sin}^2 \Theta}{2} \biggr ) 
= \frac{1}{4} (1 \pm \cos \Theta)^2 \; . 
\end{eqnarray}

With these we see that for each choice of $\Theta$ it is always possible to choose physical values 
for $f^+, \, f^-$ and $f^0$ and thus open the possibility of simplifying calculations for this example by replacing 
its spin configuration with one from Example 6.  

This replacement is easily generalised to the situation where $\Theta$ is distributed over some range. 
In that  case $P_{\rm loc} = \langle \cos \Theta   \rangle_{\Theta} $ where $ \langle  \rangle_{\Theta}$ denotes 
the average over $\Theta$. The tensor $T_{\rm loc}$ is also obtained by averaging over $\Theta$ so that
$$T_{\rm loc} =
 - \frac{3}{2} \, \biggl( \frac{ \langle {\sin}^2 \Theta   \rangle_{\Theta} }{2} - \frac{1}{3} \biggr) \, T^{\rm I} \; .$$
Clearly, $f^+, \, f^-$ and $f^0$ are now obtained by replacing the functions of $\Theta$ in 
(3.24) and (3.26) by their averages. In other words quite general conical  mixed states constructed here can be replaced by 
a configuration from Example 6. 

\vspace{2mm}

\noindent
{\bf {\small Case 2 } }

Next, I note, by comparing the ${\mathfrak T}_{\rm eq}$ of Examples 1 and 2, 
that for mixed states where $\hat q \cdot {\mathfrak J}$ has 0 as the eigenvalue, 
$${\mathfrak T}_{\rm eq} = 3 \, \biggl( \frac{ {\sin}^2 \Theta}{2} - \frac{1}{3} \biggr).$$
As required ${\mathfrak T}_{\rm eq} = -1$ when $\Theta = 0$ or $\pi$. In any case $P_{\rm loc} = 0$

In this case the corresponding configuration for  Example 6
has 
\begin{eqnarray}
f^0 = {\cos}^2 \Theta \; ,
\end{eqnarray}
and 
\begin{eqnarray}
f^+ =  f^- = \frac{ {\sin}^2 \Theta}{2} \; .
\end{eqnarray}
If $\Theta$ is distributed over some range, $f^+, \, f^-$ and $f^0$ are obtained by replacing 
${\sin}^2 \Theta$ by its average in (3.27), and (3.28).

\vspace {2mm}

\noindent
{\bf {\small Case 3} }

Finally, for a mixed state comprising fractions $F^1$ and $F^2$ of Cases 1 and 2 respectively,  
the $f^+, \, f^-$ and $f^0$ corresponding to Example 6 are obtained by averaging the
$f^+, \, f^-$ and $f^0$ from each case with weightings $F^1$ and $F^2$.

\section{Adiabatic invariants based on the ISF and ITF}
\setcounter{equation}{0}

{\color {black}As mentioned earlier, for a pure state with spin expectation value $\vec S$, 
and under appropriate conditions \cite{hde2006},
the scalar product
$I_{\rm sn} = \vec S \cdot \hat n$ is an adiabatic invariant along a trajectory, i.e., it does not change as  a parameter such
as the reference  energy is slowly varied. 
So if, for example, spins start in the eigenstates considered in Examples 1  and 5 in Section 3.3,
they  stay in those eigenstates. This is reminiscent of the Adiabatic Theorem of Quantum Mechanics \cite{Messiah}. 
For these initial states
the initial $T_{\rm loc}$ is proportional to the ITF. Then since with adiabatic variation a spin stays in its eigenstate, 
the $T_{\rm loc}$ remains  proportional to the ITF evaluated at the new parameters, with the same 
${\mathfrak T}_{\rm eq}$, namely $1/2$ in this case. On the other hand if parameters are 
changed quickly enough, $I_{\rm sn}$ changes along trajectories and I expect that $T_{\rm loc}$ ceases to be proportional to the ITF.
The variation of $I_{\rm sn}$  provides a measure of the deviation from 
adiabaticity and of course the scalar product $I_{\rm sn}$ is invariant under a rotation of the coordinate system.

I now put this picture on a quantitative  basis by defining a second
adiabatic invariant, namely  a ``scalar product'' $I_{\rm tt} := {\rm Tr}(T_{\rm loc} ~T^{\rm I})$ between the ITF and $T_{\rm loc}$ at each $(u; s)$ \cite{mv2009}. For $I_{\rm sn}$ the vector $\hat n$ acts as the reference. For $I_{\rm tt}$ the tensor $T^{\rm I}$ provides the 
reference. Since a trace is invariant under a similarity transformation, $I_{\rm tt}$ like
$I_{\rm sn}$, is  constant along trajectories when the parameters do not change. Then with
$T_{\rm loc} = {\mathfrak T}_{\rm eq} T^{\rm I}$  and the initial states, say,  of Examples  1  and 5,
$I_{\rm tt} = {\mathfrak T}_{\rm eq}$ remains at $1/2$.
So the deviation of $I_{\rm tt}$  from $1/2$ also provides a measure of the deviation from  adiabaticity.

The calculation of the variation of $I_{\rm tt}$ is straightforward.
Consider a beam with $T_{\rm loc} = {\mathfrak T}_{\rm eq} T^{\rm I}$ so that $I_{\rm tt} = {\mathfrak T}_{\rm eq}$ 
and so that on a trajectory starting at $(u^{\rm i}; s^{\rm i})$, 
$T_{\rm loc} (u^{\rm i}; s^{\rm i}) = {\mathfrak T}_{\rm eq} T^{\rm I}  (u^{\rm i}; s^{\rm i}) $. After a parameter $p$ has been varied
from the starting value $p^{\rm i}$ to a final value $p^{\rm f}$ the particle is at  $(u^{\rm f}; s^{\rm f})$
and $T_{\rm loc} (u^{\rm i}; s^{\rm i})$ has evolved to 
$R(u^{\rm i}; s^ {\rm f},s^ {\rm i} )  T_{\rm loc} (u^{\rm i}; s^{\rm i}) R^{\TR}(u^{\rm i}; s^ {\rm f},s^ {\rm i} ) $.
For brevity in the following, I will use the superscripts i and f to indicate
that objects are to be evaluated at $(u^{\rm i}; s^{\rm i}) $ and
$(u^{\rm f}; s^{\rm f}) $ respectively and I define
%${\hat m} := r~{\hat n}^{\rm i}$ 
%with $r := R(u^{\rm i}; s^ {\rm f},s^ {\rm i} ) $
{\footnote { This vector, ${\hat m}$, has no connection with the vector, ${\hat m}$, appearing in the SLIM formalism \cite{handbookbr}. }}
${\hat m} := r~{\hat n}^{\rm i}$ with $r := R(u^{\rm i}; s^ {\rm f},s^ {\rm i} ) $.
Then 
\begin{eqnarray}
I_{\rm tt}^{\rm f} &=&  {\rm Tr}(r ~ T_{\rm loc}^{\rm i} ~r{\TR} ~(T^{\rm I})^{\rm f}  )
= {\mathfrak T}_{\rm eq} {\rm Tr}(r ~ (T^{\rm I})^{\rm i} ~r{\TR} ~(T^{\rm I})^{\rm f}  )  \nonumber \\
&=& 
\sqrt{\frac{3}{2}}
{\mathfrak T}_{\rm eq} {\rm Tr}(r ~{\hat n}^{\rm i} ({\hat n}^{\rm i}) \TR ~r^{\TR}~ (T^{\rm I})^{\rm f} ) = 
\sqrt{\frac{3}{2}}
{\mathfrak T}_{\rm eq} {\rm Tr}({\hat m} {\hat m}\TR  (T^{\rm I})^{\rm f}   ) \nonumber \\
&=&  
\sqrt{\frac{3}{2}}
{\mathfrak T}_{\rm eq}{\hat m}\TR (T^{\rm I})^{\rm f} {\hat m} \nonumber \\
&=& {\frac{3}{2} } {\mathfrak T}_{\rm eq} 
\left ( {\hat m}\TR {\hat n}^{\rm f} ({\hat n}^{\rm f}) \TR {\hat m}  - \frac{1}{3} \right ) \nonumber \\
&=& {\frac{3}{2} } {\mathfrak T}_{\rm eq} 
\left ( ({\hat m}\cdot {\hat n}^{\rm f } )^2  - \frac{1}{3} \right ) 
%= \frac{3}{2}  {\mathfrak T}_{\rm eq} \left ( I_{\rm mn}^2  - \frac{1}{3} \right ) \nonumber \\
=  {\mathfrak T}_{\rm eq} 
\left ( {\frac{3}{2} }{\cos}^2 {\theta_{\rm mn}} - \frac{1}{2} \right )   \nonumber \\ 
&=& I_{\rm tt}^{\rm i} \left ( {\frac{3}{2} }{\cos}^2 {\theta_{\rm mn}} - \frac{1}{2} \right ) \; ,
\label{eq:4.1}
\end{eqnarray}
where $\theta_{\rm mn}$ is the angle 
between ${\hat m}$ and ${\hat n}^{\rm f }$.

Thus $I_{\rm tt}^{\rm f}/I_{\rm tt}^{\rm i} \le 1$ and the change in $I_{\rm tt}$ depends on just one angle
irrespective of the mixture of pure states leading to the ${\mathfrak T}_{\rm eq}$.

Like $I_{\rm sn}$, $I_{\rm tt}$ is also invariant under a rotation of the coordinate system.  
This can be exploited for calculating $I_{\rm tt}$ in another way,
for example in the reference frames of the IFF which were already exploited in Example 7 in Section 3.3
\footnote{or alternatively  in the reference frames of the field of coordinate systems formed by the principal 
axes of the ITF \cite{bkv2009}.}. 
Then as before, and in obvious notation, 
 ${\hat n}^{\rm IFF}$ has the components $(0,~1,~0)$ and  
~$(T^{\rm I})^{\rm IFF}$ is diagonal with  diagonal elements (eigenvalues)  
$(T_{11}^{\rm I})^{\rm IFF}   = (T_{33}^{\rm I})^{\rm IFF}  = - (T_{22}^{\rm I})^{\rm IFF}/2 = - 1/\sqrt 6$.
So since $T_{\rm loc}$ is traceless, $I_{\rm tt}^{\rm f} = \sqrt{3/2} ~ T_{{\rm loc},22}^{\rm f,IFF}$.
Thus  we have an interpretation of $I_{\rm tt}$ --- it simply describes $T_{{\rm loc},22}^{\rm IFF}$.

With (\ref{eq:4.1}), 
\begin{eqnarray}
T_{{\rm loc},22}^{\rm f,IFF} = T_{{\rm loc},22}^{\rm i,IFF} \left ( {\frac{3}{2} }{\cos}^2 {\theta_{\rm mn}} - \frac{1}{2} \right ) \; .
\label{eq:4.2}
\end{eqnarray}

I now note that $${\cos} \; {\theta_{\rm mn}} = {\hat m} \cdot {\hat n}^{\rm f} = {\hat m}^{\rm IFF} \cdot {\hat n}^{\rm f,IFF}
~{\rm with}~ {\hat m}^{\rm IFF} = r^{\rm IFF} {\hat n}^{\rm i,IFF} ~.$$

\vspace{3mm}

Then if ${\cos} \; {\theta_{\rm mn}} = 1$, ${\hat m}^{\rm IFF} =  {\hat n}^{\rm f,IFF}$ so that in this case  $r^{\rm IFF}$ can only represent a rotation around 
${\hat n}^{\rm i,IFF}$. Therefore with an initially diagonal  $T_{{\rm loc}}^{\rm i,IFF} = {\mathfrak T}_{\rm eq} (T^{\rm I})^{\rm i,IFF}$,
we find $T_{{\rm loc}}^{\rm f,IFF} = r^{\rm IFF} T_{{\rm loc}}^{\rm i,IFF} (r^{\rm IFF})\TR = T_{{\rm loc}}^{\rm i,IFF} 
= {\mathfrak T}_{\rm eq} (T^{\rm I})^{\rm i,IFF} = {\mathfrak T}_{\rm eq} (T^{\rm I})^{\rm f,IFF} $.
Moreover,  if  ${\cos}\;{\theta_{\rm mn}} \ne 1$,  $T_{{\rm loc}}^{\rm f,IFF} = r^{\rm IFF} T_{{\rm loc}}^{\rm i,IFF} (r^{\rm IFF})\TR$
will not be diagonal so that $T_{{\rm loc}}^{\rm f}$ will not be proportional to $(T^{\rm I})^{\rm f}$.

The vector ${\hat m} = r~{\hat n}^{\rm i}$ behaves just like a notional normalised spin starting 
parallel to $ {\hat n}$ so that 
these calculations vindicate the predictions from the heuristic discussion about the invariance of $T_{\rm loc}$
in the first paragraph of this section but without choosing special initial spin states.  
It is also now clear that  $I_{\rm tt}^{\rm f}$ does not add to the information already contained in 
the $I_{\rm sn}^{\rm f}$.
 
If, after the variation to $p^{\rm f}$, $p$ then remains fixed while the particle continues along its trajectory, 
a $T_{\rm loc}$ which is not proportional to  $T^{\rm I}$ remains non-invariant but $I_{\rm tt}$  does not change.
In other words, the fact that $I_{\rm tt}$ is constant along trajectories does not mean that $T_{\rm loc}$ is proportional to the ITF.    
If an initial  field $T_{\rm loc}(u; s)$  is proportional to the ITF and the parameters are not varied, 
${\cos}\;{\theta_{\rm mn}}$ on each trajectory will remain at the value 1 in agreement with the fact that $T_{\rm loc}(u; s)$ should  remain 
proportional to the ITF.
Obviously, if $\theta_{\rm mn}$ is the same on all trajectories $I_{\rm tt}$ is the same on 
all trajectories.

At low beam energy in a perfectly aligned simple flat storage ring with no solenoids and standard transverse emittances, where $\nu_0$ is far from an integer 
and where the ADST is far from a spin-orbit resonance, 
${\hat n}_0 (s)$ will be vertical and 
$\hat n (u; s)$ will be essentially parallel to ${\hat n}_0 (s)$ at each $s$. If a beam is injected 
with polarisation parallel to ${\hat n}_0 $  at the injection point with $\vec P(u; s) = P_{\rm inj}\;{\hat n}_0 $, 
and assuming that we can model the system as in Example 6,
then $T_{\rm loc}$ will be proportional to $T^{\rm I}$ and the same at all $(u; s)$.
Next, if the beam is accelerated in a way that ensures that the spins stay in their eigenstates 
$|n^{+}\rangle$,  $|n^{-}\rangle$  and $|n^{0}\rangle$ (so that $f^+, \, f^-$ and $f^0$ do not change) while their ISF vectors $\hat n$
might spread out
away from ${\hat n}_0$, the final
equilibrium polarisation distribution ${\vec P}_{\rm loc}(u; s) = P_{\rm inj} \; {\hat n}(u; s) $ will be established.
Moreover  the final  
$T_{\rm loc}$ will still be proportional to $T^{\rm I}$ with the same ${\mathfrak T}_{\rm eq}$. This is how an
EDMF can be established at high energy in this case.

As mentioned earlier, there is another convention for the degree of tensor polarization. This is  useful when the 
vector polarisation is vertical and results from a mixture of pure spins states with 
expectation values $\langle {\hat s}_2 \rangle$  equal to $+1, -1$ and $0$. Here, the degree of tensor polarization 
is defined to be 
the so-called alignment, $\cal A$, given by ${\cal A} := 1 - 3 f^{2,0}$ where $f^{2,0}$ is the fraction of particles with $\langle {\hat s}_2 \rangle= 0$. For this there is no need of  the concepts of ${\hat n}_0$ and $\hat n$.
%The alignment ranges from  $-2$ to $+1$ as $f^0$ ranges from  $1$ to $0$.
It is straightforward to show that $T_{22} =  1/\sqrt 6 \; \cal A $.  See \cite{Mor2003} and 
\cite[pages 52-53]{el2001}. Together with $P_2$, knowledge of the alignment is 
frequently needed  for the analysis of the results of scattering experiments at low energies involving deuterons. 
See, for example,  the discussion surrounding 
\cite[eq. 3.1.72]{el2001}. There, only $P_2$ and $T_{22}$ appear in the density 
matrix and it is diagonal. Thus other components of the tensor are not needed.
See \cite{pgs2011, conzett1994} too.  

The component $T_{22}$ of the tensor provides a useful window into the behaviour of the 
tensor part of the spin dynamics of deuterons in storage rings at low energy.
Consider a perfectly aligned simple flat storage ring at low beam energy with standard  transverse emittances and no solenoids where $\nu_0$ is far from an integer so that ${\hat n }_0 (s)$ is vertical.
If the ADST is far from a spin-orbit resonance 
then  $\hat n (u; s)$ will be essentially vertical too at each $(u; s)$.
If the vector polarisation is $\vec P(u; s) = P_{\rm inj}\;{\hat n}_0 $ and the mixed state can be modelled as in Example 6, then the EDMF is diagonal and
${\sqrt 6} T_{{\rm loc},22}$ 
is simply the local alignment ${\cal A}_{\rm loc}$ \cite[pages 52-53]{el2001} introduced earlier.
Moreover, from (3.21), ${\mathfrak T}_{\rm eq} = 1/2 (1 - 3 f^{2,0} )$ so that the two definitions of the degree of tensor polarisation differ just by a factor of two for this configuration. 
Next, if, for example, the vertical betatron tune or the frequency of a radio-frequency dipole is varied non-adiabatically
through, say,  a state of spin-orbit resonance to a state far from resonance, 
$\hat n$ will again be very close to the vertical at all $(u; s)$ at the end of the variation. 
However, ${\cos}\;{\theta}_{\rm mn}$ for each particle will   
in general differ from its initial value of 1. 
Then, as we have seen, 
the final $T_{\rm loc}$ is not proportional to the $T^{\rm I}$ and at each $(u; s)$
${T_{{\rm loc},22}^{\rm f}} = {T_{{\rm loc},22}^{\rm i}} ( {\frac{3}{2} } { {\cos}^2} {\theta}_{\rm mn} - \frac{1}{2} ) $.
If the Froissart-Stora formalism applies \cite{fs60}, $\theta_{\rm mn}$ will be the same for all trajectories.
For related experimental work with a radio-frequency dipole see \cite{Mor2003,Mor2005}.
In particular, if the parameter variation is, in fact, adiabatic,  ${\cos}\;\theta_{\rm mn} = 1$ 
so that ${T_{{\rm loc},22}^{\rm f}} = {T_{{\rm loc},22}^{\rm i}}$.

However, at high energy and/or close to a spin-orbit resonance, $\hat n (u; s)$ can be tilted away from 
${\hat n}_0(s)$. Then although $T_{\rm loc}$ can be proportional to $T^{\rm I}$, the $T^{\rm I}$ is not diagonal
and $T_{{\rm loc},22}$ is  not, on its own, sufficient for analysing scattering experiments.  
Nevertheless, insights are available via numerical simulation. For example the system can be viewed in 
a reference frame of the IFF or w.r.t. principal axes. 
Then if the vector polarisation at $(u; s)$ 
is parallel to ${\hat n} (u; s)$, the density matrix is diagonal and depends just on $|P_{\rm loc } (u; s)|$ and on the $T_{{\rm loc},22}$
viewed in that frame. With this, the analysis of the effect of a variation of a parameter can be carried through
in this frame (or w.r.t. the principal axes) in the same way as in the discussion leading to (\ref{eq:4.2}).
An example of working in a reference frame of the IFF for protons is given in \cite{spin2010}. See \cite{hv2004} too.

\section{Other parametrisations}
Once the Cartesian version of the ITF has been established, the five
components of the corresponding invariant spherical
polarisation--tensor field follow trivially using the relations in
\cite[eqs. 3.1.66 - 67]{el2001}. See \cite{werle1966,pgs2011,conzett1994,MC1970}
too.  Thus it is simple to find the invariant spherical
polarisation--tensor field for, for example,  the single resonance model.  In analogy
with the expansion for spin-1/2 particles in (\ref{eq:1.1}), the
density matrix for spin-1 particles can also be expanded in terms of
the generators of the group SU(3) and the so--called coherence (or
Bloch) vector \cite{byrd,kimura}.  Other matrix bases and the
corresponding Bloch vectors can be used too: see  \cite{bertlmann2008}
and the Appendix too.
However, the expansion in terms of the polarisation and the Cartesian
tensor seems to be the most convenient for calculating and discussing
the EDMF.  Moreover, although 
not all Bloch vectors inside their so--called Bloch hyper--spheres lead to
admissible density matrices \cite{bertlmann2008}, this matter does not arise for density matrices based on the ITFs
of (\ref{eq:7}).

\section{The Bloch equations}
\setcounter{equation}{0} 
In the previous sections, the orbital motion is deterministic and governed just by a
Hamiltonian so that the density in phase space is conserved along a
trajectory and one can work with spin density matrices instead of
spin--orbit density matrices. However, if the particles are subject to
noise and damping, the evolution of the density of particles in phase
space is more complicated and spin--orbit density matrices and their
Wigner functions come into play \cite{mont98,dk75}. Alternatively, we
can work with the classical phase space density and the so--called
polarisation density as in \cite{dbkh98,rbg99}.  The effect of noise
and damping on the polarisation tensor can be studied in an analogous
way. So 
I round off this paper on the polarisation tensor by showing how to do this.
In order to avoid undue
repetition I assume that the reader is familiar with the concepts in
\cite{dbkh98,rbg99}. The final result of this section is perhaps no more 
than a curiosity but 
it would be remiss not to include it here while the ITF is being presented
in such detail.

Charged particles suffering deflection in magnetic fields emit
synchrotron radiation and, as is well known, this has important
consequences for the properties of electron (positron)  beams in storage rings.
For example, the stochastic nature of photon emission together 
with damping mechanisms causes the phase space density $W_{\rm
  orb}(u; s)$ to reach equilibrium, i.e., to become 1--turn periodic: $W_{\rm orb}(u; s +C) = W_{\rm
  orb}(u; s)$ \cite{wiedemann}. 
It often suffices to approximate the effects of the photon emission as
a Gaussian white noise process so that
if interparticle forces can be ignored,
 the evolution of $ W_{\rm orb}$
can be described in terms of the Fokker--Planck equation
\cite{wiedemann,
  risken,gardiner,vankampen,mr8362,bhmr91,jowett,prr,khdecoh,
  handbookemr,tzenov}. For this account it is sufficient to write the
Fokker--Planck equation somewhat symbolically as in \cite{dbkh98} in
the form
\begin{eqnarray}
\frac{\partial W_{\rm orb}} {\partial s}   =
       {\cal L}_{_{\rm FP,orb}} \; W_{\rm orb} \; ,
\label{eq:13}
\end{eqnarray}
where the orbital Fokker--Planck operator can be decomposed into the form:
\begin{eqnarray}
{\cal L}_{_{\rm FP,orb}} = {\cal L}_{\rm ham} + {\cal L}_{0} + {\cal L}_{1} + 
{\cal L}_{2} \; , 
\label{eq:14}
\end{eqnarray}
whereby  
${\cal L}_{0},~ {\cal L}_{1},~{\cal L}_{2}$ are terms due to damping and noise
containing respectively zeroth, first and second order derivatives w.r.t. the components of
$u$. The term  
${\cal L}_{\rm ham}$ is the Poisson bracket 
$\lbrace{{H}}_{\rm orb} ,W_{\rm orb} \rbrace$
of $W_{\rm orb}$ with the orbital Hamiltonian.
Detailed forms for ${\cal L}_0$, ${\cal L}_1$ and ${\cal L}_2$  can be found in
\cite{bhmr91,jowett,prr}
but are not important for the argument that follows. 
We normalise $W_{\rm orb}$
to unity: $ \int d^6 u ~ W_{\rm orb}(u; s) = 1$.

I have introduced the Fokker--Planck equation by mentioning
synchrotron radiation but the Fokker--Planck equation can be applied
when the particles are subject to other sources of noise and damping
by adopting appropriate forms for ${\cal L}_{0},~ {\cal L}_{1}$ and
${\cal L}_{2}$. Of course, except at very large energies, synchrotron
radiation is irrelevant for particles other than electrons (positrons).
But as in the case of synchrotron radiation, noise can lead to
irreversible loss of vector polarisation \cite{mont98}, and even when
$W_{\rm orb}$ has reached periodicity. The chief mechanism is simple:
the noise causes random perturbations to the trajectories and thereby
causes random perturbations in $\vec \Omega$ in the non--uniform fields
of the quadrupoles. 

Since the local vector polarisation ${\vec P}_{\rm
  loc}(u; s)$  is not a density, its evolution cannot be described by a
Fokker--Planck equation.  However, as explained in \cite{dbkh98,rbg99},
if the noise has no {\em direct} effect on spins, the
vector--polarisation density $\vec {\cal P}(u; s) := W_{\rm orb}(u; s) {\vec
  P}_{\rm loc}(u; s)$ evolves according to the equation
\begin{eqnarray}
 \frac{\partial \vec{\cal P}}{\partial s}~ =~ 
             {\cal L}_{_{\rm FP,orb}}\vec{\cal P} 
        +   \vec{\Omega}  \times   \vec{\cal P}  \; ,
\label{eq:15}
\end{eqnarray}
where the Poisson bracket ${\cal L}_{\rm ham}$ is now
$\lbrace{{H}}_{\rm orb}, \vec {\cal P} \rbrace$.  Thus, if ${\cal
  L}_{_{\rm FP,orb}}$ is known, we have an immediate, succinct,
classical encapsulation of the way in which ${\vec P}_{\rm loc}$ is
modified both by precession and by the effect of noise and damping on
the mixture of spin states at each $(u; s)$.  We call (\ref{eq:15}) the
Bloch equation for the vector--polarisation density to reflect the analogy with the equations for magnetisation
in magnetised solids \cite{FBloch}{\footnote {This equation has nothing directly to do with the Bloch vectors mentioned earlier.}}.
Although (\ref{eq:15}) was derived with spin-1/2
particles in mind, it applies to spin-1 particles too.  
The vector--polarisation density is proportional to the density in
phase space of spin angular momentum.
To obtain ${\vec P}_{\rm loc}(u; s)$,
(\ref{eq:13}) and (\ref{eq:15}) should be solved in parallel and then
$\vec {\cal P}(u; s) / W_{\rm orb}(u; s)$ should be calculated.

If ${\cal L}_{0},~ {\cal L}_{1}$ and ${\cal L}_{2}$ vanish, leaving just the Poisson
bracket ${\cal L}_{\rm ham}$, then  (\ref{eq:15})   
reduces to the T--BMT equation for $\vec {\cal P}$ along a trajectory, 
just as expected when we recall that in this case
$W_{\rm orb}$ is preserved along the trajectory.
Then, if ${\vec P}_{\rm loc}(u; s)$ is parallel to ${\hat n}(u; s)$  at each point in phase space with 
${\vec P}_{\rm loc}(J,\phi; s) = {P}_{\rm eq}(J) \, \hat
n(J,\phi; s)$,
and $W_{\rm orb}$ is in equilibrium, the polarisation of the beam will be independent of $s$.
On the other hand if ${\vec P}_{\rm loc}(u; s)$ is not parallel to ${\hat n}(u; s)$, 
the polarisation of the beam will oscillate as illustrated in \cite[p.72]{gh2006}.
If noise and damping are included, the polarisation of the beam will die away in the long term 
\cite{mont98,bhmr91,dbkh2015}.

In order to study the evolution of the Cartesian polarisation tensor I
introduce the polarisation-tensor density
${\cal T}(u; s) := W_{\rm orb}(u; s) T_{\rm loc}(u; s)$.
Then it can be shown that the Bloch--like equation for ${\cal T}$ is
\begin{eqnarray}
 \frac{\partial {\cal T}}{\partial s}~ =~ 
             {\cal L}_{_{\rm FP,orb}}{\cal T} 
        +   [\, {\tilde {\Omega}}, \, {\cal T}]  \; ,
\label{eq:16}
\end{eqnarray}
where the commutator derives from (\ref{eq:4.1}) and 
where the Poisson bracket ${\cal L}_{\rm ham}$ is  
$\lbrace{{H}}_{\rm orb}, {\cal T }  \rbrace$.
Given the previous discussion, (\ref{eq:16}) is already expected on purely heuristic grounds. 
In fact (\ref{eq:16}) is valid for an arbitrary tensor obeying (\ref{eq:4.1}) but, in particular,
for the traceless symmetric tensor $T_{loc}$. 
To obtain ${T}_{\rm loc}(u; s)$,
(\ref{eq:13}) and (\ref{eq:16}) should be solved in parallel and then
${\cal T}(u; s) / W_{\rm orb}(u; s)$ should be calculated.

Since, by definition, ${\cal T}$ is symmetric and traceless at all $(u; s)$,
the right-hand side of (\ref{eq:16}) is traceless and symmetric too.
Therefore I expect that (\ref{eq:16}) preserves the
symmetry and tracelessness of ${T}_{\rm loc}$.  
In any case, this should happen since the form (\ref{eq:2.3.1}) must be preserved in the presence of noise and damping.
In the absence of a polarising
mechanism, $P_{\rm loc}$ falls  to zero in beams of spin-1/2
particles subject to noise and damping \cite{mont98,bhmr91,dbkh2015,handbookbr}. This will also happen for
spin-1 particles and 
I expect that ${\mathfrak T}_{\rm loc}$ will
fall to zero {\color {black} as the spin states become fully mixed.}
Moreover,  with noise and damping $P_{\rm loc}(u; s)$ aligns itself almost parallel to ${\hat n}(u; s)$ \cite{dbkh2015} during the depolarisation. 
Then I expect that ${T}_{\rm loc}(u; s)$ becomes almost proportional to ${T^I}(u; s)$ at the same time.

Note that the relationships between (\ref{eq:13}) and (\ref{eq:15})
and between (\ref{eq:13}) and (\ref{eq:16}) survive if ${\cal
  L}_{_{\rm FP,orb}}$ is replaced by any physically admissible
transport operator ${\cal K}_{_{\rm orb}}$ --- which could even
contain derivatives beyond second order.

\section{Summary}
I have proposed a definition of an invariant rank--2 Cartesian
polarisation--tensor field (ITF) for spin-1 particles, and a procedure for
calculating it numerically by stroboscopic averaging or analytically
once the ISF is known.  The ISF and ITF provide  ``chassis'' on
which to ``hang'' equilibrium vector and tensor spin distributions for spin-1 particles
on a phase--space torus and I have adopted the two fields to construct
equilibrium spin density--matrix fields which depend on just two
parameters, the degree of equilibrium vector polarisation $P_{\rm
  eq}(J)$ and the degree of equilibrium tensor polarisation
${{\mathfrak T}_{\rm eq}}(J)$.  I have also pointed out that the EDMF
for spin-1 particles can always be diagonalised by a rotation of the
coordinate system when the ansatz (\ref{eq:7}) for the ITF is valid and I have
shown with examples how that ansatz can accommodate  a typical spin-1 tensor in a natural way.
Moreover, I have identified an adiabatic invariant associated with the ITF. 
Of course this work is mainly relevant for deuterons but  
from the arguments in this paper, it is clear that an EDMF and an 
accompanying invariant tensor field could be defined for spin-3/2 particles
too. Of course, such objects would have no practical use owing to the 
small lifetimes of those particles.
 
Finally, I have extended earlier work to include the effects of noise
and damping on the polarisation tensor and I have provided an evolution equation 
for the polarisation-tensor density.

A follow-up paper \cite{bkv2009} will discuss the uniqueness of the ansatz in (\ref{eq:7}).

\section*{Acknowledgments}
I thank D.T. Abell, J.A. Ellison, K. Heinemann, A. Kling, E. Leader, V. Morozov, E. Stephenson  and M. Vogt
for valuable comments and suggestions and/or fruitful collaboration.  
I also thank O. Giraud for discussions on the connection between this formalism and that in \cite{giraud2008}. 

\addcontentsline{toc}{section}{Appendix }
\section*{Appendix}
In this appendix I give a heuristic and pedagogical demonstration of how to arrive at the representation
(\ref{eq:2.3.1}) for the spin-1 density matrix.

We need to write $\rho$ as a linear combination of $3 \times 3$
matrices.  There is a variety of possibilities
\cite{byrd,kimura,bertlmann2008} but we wish to use the matrices
${\mathfrak J}$. Since eight real parameters are needed, an expansion
of the form $\rho = \sum_{i= 1-3} U_{i} \, {\mathfrak J}_i$ with complex $U_{i}$ does not
suffice. So I try an expansion of the form 
\begin{equation}
\rho = \sum_{i,j} U_{ij} \,{\mathfrak J}_j {\mathfrak J}_i  \; ,
\tag{A.1.1}
\end{equation}
with complex $U_{ij}$. Then we have space for the required eight
independent real parameters but their identity is not immediately evident among the eighteen 
real parameters defining the $U_{ij}$.

To come further I recall that  ${\rm Tr}({\mathfrak J}_i {\mathfrak J}_j) = 2 \delta_{ij}$.
Then ${\rm Tr}(U) = 1/2$ since ${\rm Tr}(\rho)= 1$.
Moreover, with the rule 
$\langle\hat s_i \, \hat s_j  \rangle = {\rm Tr}(\rho {\mathfrak J}_i {\mathfrak J}_j )$ 
I then find 
\begin{equation}
U_{ij} = \langle\hat s_i \hat s_j \rangle - \frac{1}{2} \delta_{ij} \; , 
\tag{A.1.2}
\end{equation}
which I can write in matrix form as 
\begin{equation}
U  = \langle\hat s \, \hat s\TR \rangle - \frac{1}{2} I \; , 
\tag{A.1.3}
\end{equation}
in an obvious notation. The matrix $ \langle\hat s \, \hat s\TR \rangle$
transforms as a Cartesian tensor. Therefore $U$ transforms as a
Cartesian tensor too. Thus $U^{\rm f} =    R ~U^{\rm i} R\TR$ in analogy
with (\ref{eq:4}) so that ($U\TR)^{\rm f} =    R ~(U\TR)^{\rm i} R\TR$. 

For the next step I write  $U$ as the sum of its symmetric and antisymmetric parts:
$U = t + a $ with $t = (U + U\TR)/2$ and $a = (U - U\TR)/2$. Both $t$ and
 $a$ are Cartesian tensors.
Furthermore, using the hermiticity of  $\rho$ it can be shown that $t$ is real and $a$ is 
pure imaginary.

Equation (A.1.1) can now be written as
\begin{equation}
\rho =  \sum_{i,j} \frac{t_{ij}}{2}({\mathfrak J}_i {\mathfrak J}_j + {\mathfrak J}_j {\mathfrak J}_i) 
 + a_{12} ({\mathfrak J}_1 {\mathfrak J}_2 - {\mathfrak J}_2 {\mathfrak J}_1)
 + a_{23} ({\mathfrak J}_2 {\mathfrak J}_3 - {\mathfrak J}_3 {\mathfrak J}_2) 
 + a_{13} ({\mathfrak J}_1 {\mathfrak J}_3 - {\mathfrak J}_3 {\mathfrak J}_1) \; ,
\tag{A.1.4}
\end{equation}
and from  the commutation relations among the matrices ${\mathfrak J}$ we obtain 
\begin{equation}
\rho = \sum_{i,j} \frac{t_{ij}}{2}({\mathfrak J}_i {\mathfrak J}_j + {\mathfrak J}_j {\mathfrak J}_i) 
 + i a_{12} {\mathfrak J}_3
 + i a_{23} {\mathfrak J}_1
 - i a_{13} {\mathfrak J}_2 \; .
\tag{A.1.5}
\end{equation}
I now define a real  3--component object $\vec v$ with elements
$\vec v_1 := i \, a_{23}, \vec v_2 := - i \, a_{13}$ and $\vec v_3 := i \,a_{12}$. Then, using the 
fact that $a$ is a Cartesian tensor, it is easily shown  that $\vec v$ transforms as an axial 3--vector. 
Thus (A.1.5) takes the form
\begin{equation}
\rho = \sum_{i,j} \frac{t_{ij}}{2}({\mathfrak J}_i {\mathfrak J}_j + {\mathfrak J}_j {\mathfrak J}_i) 
 + {\vec v} \cdot \vec {\mathfrak J} \; .
\tag{A.1.6}
\end{equation}
With the definition $C = t/2$ so that ${\rm Tr}(C)= 1/4$, I rewrite (A.1.6) as 
\begin{equation}
\rho = \sum_{i,j} C_{ij}({\mathfrak J}_i {\mathfrak J}_j 
+ {\mathfrak J}_j {\mathfrak J}_i) 
 + {\vec v} \cdot \vec {\mathfrak J} \; .
\tag{A.1.7}
\end{equation}
This, in turn, can be rearranged in terms of the traceless tensor $C - \frac{1}{12} I$ to give 
\begin{equation}
\rho = \frac{1}{3} I + \sum_{i,j} (C_{ij} - \frac{1}{12} \delta_{ij}) ({\mathfrak J}_i {\mathfrak J}_j 
+ {\mathfrak J}_j {\mathfrak J}_i) 
 + {\vec v} \cdot \vec {\mathfrak J} \; .
\tag{A.1.8}
\end{equation}
Then by making the identifications $\vec P = {\rm Tr}(\rho \vec {\mathfrak J} ) = 2 \vec v$ and
$\frac{1}{\sqrt{6}} T = C - \frac{1}{12} I$ I arrive at (\ref{eq:2.3.1}).

The passage from (A.1.1) to (A.1.8) is a demonstration of how to usefully decompose the tensor $U$
into its scalar, vector and irreducible tensor parts \cite{bell75}.
Moreover, by explicitly isolating the scalar term, $\frac{1}{3} I$, in (A.1.8) we obtain a transparent form 
for $\rho$  when the three spin substates are equally populated: $\vec P$ and $T$ 
must vanish. 

The parametrisation (A.1.8) contains nine real parameters, but of course,
only eight of them are independent owing to the condition ${\rm Tr}(C) =
1/4$. However, we can also begin with a parametrisation based on the
matrices ${\mathfrak J}$ but with just eight parameters. For this 
we recall  that the density matrix can be written in the form:
\begin{equation}
\rho = \frac{1}{3} I +\sum_{k =1-8} \lambda_k {\cal O}_k \; ,
\tag{A.1.9}
\end{equation}
where the $3 \times 3$ matrices $\cal O$ are traceless and hermitian 
and ${\rm Tr}({\cal O}_k {\cal O}_l) = \alpha_k \delta_{kl}$ so that 
together with $\frac{1}{3} I$ they comprise an orthogonal basis.  
The coefficients $\alpha$ are real and the components, $\lambda_k = {\rm Tr}(\rho \, {{\cal O}_k} )/\alpha_k$, of the Bloch vector are real 
and mutually independent.  

Then we can, for example, choose the  eight assignments:

\begin{eqnarray}
\qquad \qquad \qquad {\cal O}_1 &=& {\mathfrak J}_1   \qquad \qquad \qquad \,
{\cal O}_2 \, = \,\,{\mathfrak J}_2 \qquad \qquad\qquad \!
{\cal O}_3 \, = \,\, {\mathfrak J}_3                      \nonumber  \\
{\cal O}_4 &=& {\mathfrak J}_1 {\mathfrak J}_2 
              + {\mathfrak J}_2 {\mathfrak J}_1 \qquad 
{\cal O}_5 \, =\, \, {\mathfrak J}_2 {\mathfrak J}_3 
              + {\mathfrak J}_3 {\mathfrak J}_2 \qquad \!
{\cal O}_6 \, = \, \, {\mathfrak J}_1 {\mathfrak J}_3 
             + {\mathfrak J}_3 {\mathfrak J}_1  \nonumber  \\
{\cal O}_7 &=& {\mathfrak J}_1 {\mathfrak J}_1 
              - {\mathfrak J}_3 {\mathfrak J}_3 \qquad 
{\cal O}_8  \, = \, \, {\mathfrak J}_2 {\mathfrak J}_2 - \frac{2}{3}  I \; . \hspace*{50 mm} (A.1.10)
\nonumber
\end{eqnarray}
%\vspace{3mm}
It is simple to show that, as required, 
$\langle \hat s_1 \, \hat s_1 +  \hat s_2 \, \hat s_2  + \hat s_3 \, \hat s_3 \rangle = 2$
independently of the coefficients $\lambda$, i.e, for any mixed state.
Of course, $\vec P$ and $T$ are simply related to the coefficients $\lambda$.
Moreover, the parametrisation (A.1.10) is convenient for using  
the equation of motion of the density matrix as in \cite[Section 1--8c]{roman}, but with the commutation 
relations for the matrices $\mathfrak J$,   
to demonstrate that  
$\vec P$ for spin--1 particles obeys the T-BMT equation. That approach also leads  
to the equation of motion  (\ref{eq:4.1}) for $T$.

\addcontentsline{toc}{section}{References}
\small \small


\begin{thebibliography}{99}
\bibitem{mv2009}
I thank M. Vogt for suggesting that there might be adiabatic invariants associated
with the ITF and that traces could be a way of quantifying them.
%
\vspace{-3mm}
\end{thebibliography}
\end{document}